\documentclass{article}

\usepackage[final]{neurips_data_2023}
\usepackage[colorinlistoftodos,prependcaption,textsize=tiny]{todonotes}

\usepackage[utf8]{inputenc} 
\usepackage[T1]{fontenc}    
\usepackage{url}            
\usepackage{booktabs}       
\usepackage{amsfonts}       
\usepackage{nicefrac}       
\usepackage{microtype}      
\usepackage{xcolor}         
\usepackage{amsmath, bm}
\usepackage{multirow}
\usepackage{graphicx}
\usepackage{threeparttable}
\usepackage{soul}
\usepackage{subcaption,wrapfig}
\definecolor{light}{rgb}{0.3, 0.3, 0.3}
\def\light#1{{\color{light}#1}}

\usepackage{hyperref}

\title{QH9: A Quantum Hamiltonian Prediction Benchmark for QM9 Molecules}

\author{%
  Haiyang Yu\thanks{Equal contribution}\\
  Texas A\&M University\\
  College Station, TX 77843 \\
  \texttt{haiyang@tamu.edu} \\
   \And
  Meng Liu\footnotemark[1]\\
  Texas A\&M University\\
  College Station, TX 77843 \\
  \texttt{mengliu@tamu.edu} \\
   \And
  Youzhi Luo\\
  Texas A\&M University\\
  College Station, TX 77843 \\
  \texttt{yzluo@tamu.edu} \\
   \And
  Alex Strasser\\
  Texas A\&M University\\
  College Station, TX 77843 \\
  \texttt{alexstrasser16410@tamu.edu} \\
   \And
  Xiaofeng Qian\thanks{Equal senior contribution}\\
  Texas A\&M University\\
  College Station, TX 77843 \\
  \texttt{feng@tamu.edu} \\
   \And
  Xiaoning Qian\footnotemark[2]\\
  Texas A\&M University\\
  College Station, TX 77843 \\
  \texttt{xqian@ece.tamu.edu} \\
   \And
  Shuiwang Ji\footnotemark[2]\\
  Texas A\&M University\\
  College Station, TX 77843 \\
  \texttt{sji@tamu.edu} 
}

\begin{document}

\maketitle
\newcommand{\red}[1]{\textcolor{red}{#1}}
\newcommand{\toauthor}[1]{\textbf{\textcolor{red}{#1}}}

\begin{abstract}
Supervised machine learning approaches have been increasingly used in accelerating electronic structure prediction as surrogates of first-principle computational methods, such as density functional theory (DFT). 
While numerous quantum chemistry datasets focus on chemical properties and atomic forces, the ability to achieve accurate and efficient prediction of the Hamiltonian matrix is highly desired, as it is the most important and fundamental physical quantity that determines the quantum states of physical systems and chemical properties. In this work, we generate a new Quantum Hamiltonian dataset, named as QH9, to provide precise Hamiltonian matrices for 999 or 2998 molecular dynamics trajectories and 130,831  stable molecular geometries, based on the QM9 dataset. By designing benchmark tasks with various molecules, we show that current machine learning models have the capacity to predict Hamiltonian matrices for arbitrary molecules. Both the QH9 dataset and the baseline models are provided to the community through an open-source benchmark, which can be highly valuable for developing machine learning methods and accelerating molecular and materials design for scientific and technological applications. Our benchmark is publicly available at \url{https://github.com/divelab/AIRS/tree/main/OpenDFT/QHBench}.
\end{abstract}

\section{Introduction}
\begin{figure}
    \centering
    \includegraphics[width=1.0\textwidth]{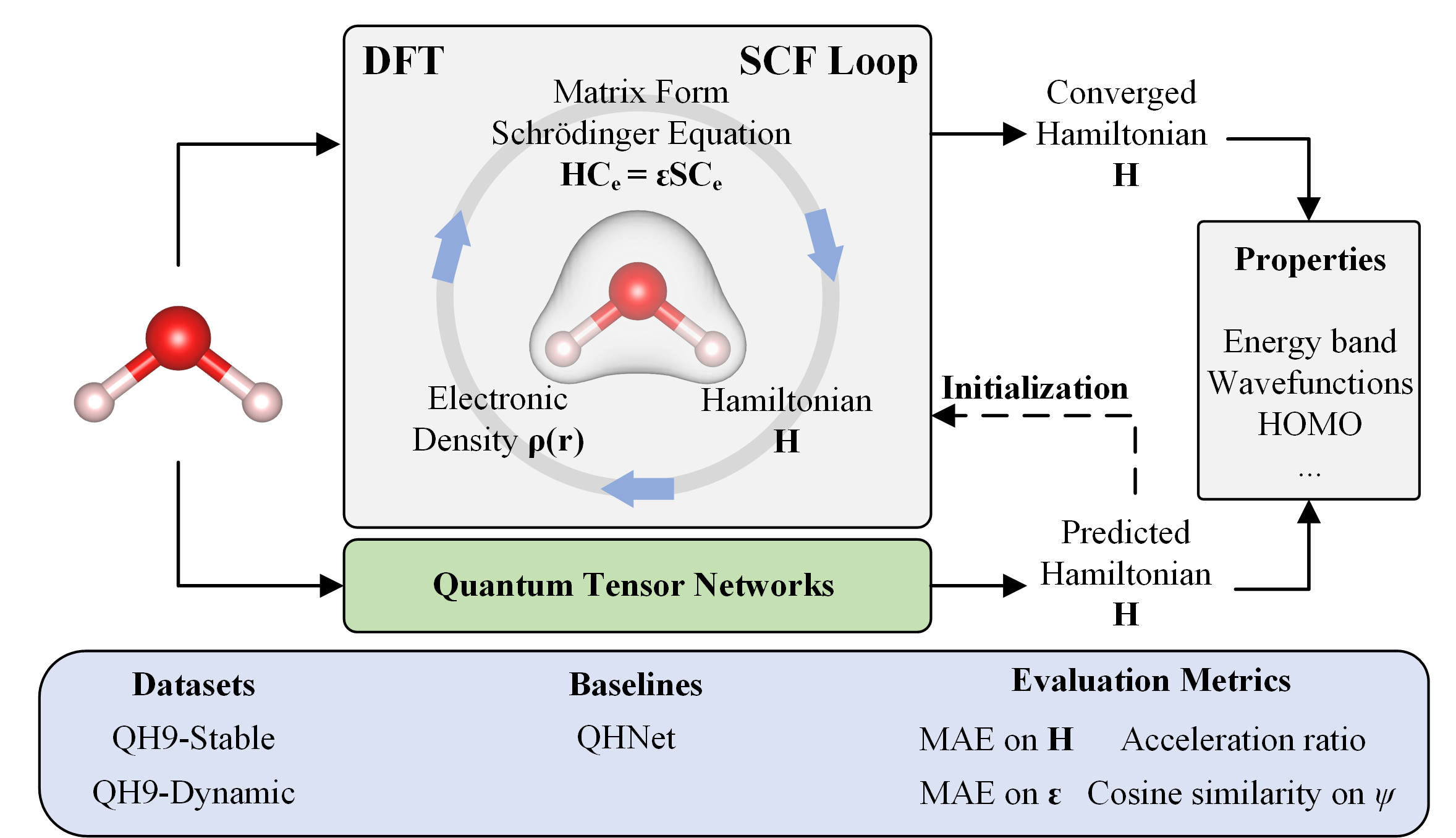}
    \caption{The target and content of the propose QH9 dataset and benchmark. Quantum tensor networks are built for predicting the target Hamiltonian matrix, facilitating the optimization loop within DFT by providing a precise approximation. 
    Within this benchmark, stable and dynamic datasets are generated for training powerful quantum tensor networks and comprehensive evaluation metrics are proposed to measure the prediction quality.}
    \label{fig:QH9}
\end{figure}

Machine learning methods have shown great potential in accelerating computations in quantum chemistry tasks \citep{zhang2023artificial, huang2023central, kirkpatrick2021pushing}. For example, a variety of invariant geometric deep learning methods have been developed to encode pairwise distances and bond angles in molecular and materials systems \citep{schutt2018schnet, gasteigerdirectional, gasteiger2021gemnet, liu2021spherical, wang2022comenet, lin2023efficient, yan2022periodic, liao2023equiformerv2, schutt2021equivariant, batzner20223, liu2021dig} to accelerate the prediction of their chemical properties as data-driven surrogate approximations. To enhance the prediction of vectorial properties, such as force fields, equivariant deep learning methods have been developed to capture permutation, translation, and rotation equivariance for equivariant property prediction \citep{satorras2021n, schutt2021equivariant, tholke2022equivariant, thomas2018tensor, batzner20223, fuchs2020se, liao2023equiformer, anderson2019cormorant, brandstetter2022geometric, batatia2022mace}. To support and facilitate the development of machine learning methods on quantum chemistry property prediction, many datasets have been generated to benchmark the respective tasks on molecular property prediction \citep{blum2009970, ruddigkeit2012enumeration, ramakrishnan2014quantum, wang2009pubchem, nakata2017pubchemqc}, catalyst prediction \citep{chanussot2021open, oc22}, and force field prediction \citep{chmiela2017machine, chmiela2023accurate}. 

In addition to these quantum chemistry prediction tasks, the quantum Hamiltonian is another significant and fundamental physical property that determines the quantum states and various materials properties~\citep{Marzari1997mlwf, Souza2001mlwf, qian2010qotransport, Marzari2012rmp, Bai2022graph}. The quantum Hamiltonian can be calculated using Density Functional Theory~(DFT)~\citep{hohenkohn, kohnsham} with a time complexity of $O(n^{3} T)$, where $n$ represents the number of electrons and $T$ denotes the number of optimization steps required to achieve convergence. 
Given the high computational complexity of the DFT algorithms, accelerating such calculations for novel molecular and materials systems becomes a desirable but challenging task.
To tackle this challenge, machine learning methods, such as quantum tensor networks \citep{li2022deep, gong2023general, schutt2019unifying, yu2023efficient, unke2021se}, provide a highly promising approach for accelerating the DFT algorithms. These networks directly predict the final Hamiltonian matrix given the input 3D geometries, resulting in significant acceleration of calculations by orders of magnitude. 

Unlike invariant chemical properties, Hamiltonian matrices obey intrinsic block-by-block matrix equivariance. This equivariance can be represented by the rotation Wigner D-Matrix, which may contain higher order rotations beyond 3D space.
In order to make physically meaningful predictions, it is important to design quantum tensor network architectures that preserve this equivariance property.
To perform systematic and in-depth study of this new task, there is a clear need to generate large-scale quantum tensor datasets and benchmarks.
Currently, existing quantum Hamiltonian datasets include the MD17 \citep{schutt2019unifying, gastegger2020deep} and mixed MD17 \citep{yu2023efficient} datasets, which consist of data for a single and four molecules, respectively. Recent public dataset NablaDFT~\citep{khrabrov2022nabladft} contains Hamiltonian matrices for molecular conformations.

To provide a realistic dataset and comprehensive evaluation, we generate a new quantum tensor dataset named QH9. This dataset contains Hamiltonian matrices for 130,831 stable molecular geometries and 999 or 2998 molecular dynamic trajectories. In order to provide comprehensive studies for quantum tensor networks, we have designed four specific tasks.
The first two tasks, QH9-stable-id and QH9-stable-ood, aim to explore the performance of the networks in both in-distribution and out-of-distribution scenarios, specifically focusing on stable molecular geometries.
The QH9-dynamic-geo task follows the setting of the mixed MD17 \citep{yu2023efficient}, containing the same molecule with different geometries in the training, validation, and test. On the other hand, the QH9-dynamic-mol task splits the trajectories based on different molecules. Based on the number of geometries within the QH9-dynamic datasets, it is divided into QH9-dynamic-100k for 999 trajectories and QH9-dynamic-300k for 2998 trajectories.
Finally, we evaluate the transferability of the trained models on molecules with larger sizes, thereby testing the models' ability to generalize beyond the training dataset.
To demonstrate the quality of the predicted Hamilton matrix, we use four metrics. These metrics are based on the Mean Absolute Error (MAE) of the predicted Hamiltonian matrix $\bm{H}$, as well as the derived properties such as orbital energies $\boldsymbol{\epsilon}$ and electronic wavefunction $\psi$. Furthermore, to evaluate the quality of the predicted Hamiltonian in accelerating DFT calculations, we calculate the DFT optimization ratio by taking the model predictions as DFT initialization. The target and content of the proposed QH9 dataset and benchmark are demonstrated in Figure \ref{fig:QH9}.

\section{Background and Related Works}
\subsection{Density Functional Theory~(DFT)}

Modeling the quantum states of physical systems is a central topic in computational quantum physics and chemistry. It aims to solve the Schr\"{o}dinger equation~\citep{schrodinger1926undulatory}, which describes the electronic states shown as 
\begin{equation}
    \hat{H} \Psi \left( \bm{r}_1, \cdots, \bm{r}_n \right) = E \Psi \left( \bm{r}_1, \cdots, \bm{r}_n \right),
    \label{eq:schrodinger}
\end{equation}
where $\Psi \left( \bm{r}_1, \cdots, \bm{r}_n \right)$ is the $n$-electronic wavefunctions and $\bm{r}$ is the 3D coordinates. Electronic eigenvalues and wavefunctions play an important role in calculating numerous crucial physical properties, including the Highest Occupied Molecular Orbital (HOMO), the Lowest Unoccupied Molecular Orbital (LUMO), charge density and many others. However, due to the exponentially expanding input Hilbert space with the number of electrons, the computational cost to directly calculate many-electronic wavefunctions is extremely high. Therefore, various methods are proposed to approximate the solutions, such as the Hartree-Fock (HF) method \citep{szabo2012modern} that approximates the wavefunction itself, or density functional theory \citep{hohenkohn} that approximates the electron density. While the HF method scales with the number of electrons $n$ as $O(n^{4} T)$, DFT scales with $O(n^{3} T)$ and therefore DFT is better suited for large-scale systems. DFT is based on the key discovery that the total energy and thus all ground-state properties of a system are uniquely determined by the ground-state electron density~\citep{kohnsham}. 

Both of these approaches divide an $n$-electron system into a set of $n$ non-interacting one-electron wavefunctions $\psi_i(\bm{r_i})$, also called molecular orbitals in molecular systems. These one-electron orbitals can then be approximated by a linear combination of basis functions $\phi_j(\bm{r})$ as
    $\psi_i(\bm{r}) = \sum_j C_{ij} \phi_j(\bm{r}).$
The basis functions can be represented in analytical forms, such as Slater-type orbitals (STOs), Gaussian-type orbitals (GTOs), or plane waves, under numerical approximations for obtaining the coefficients matrix $\bm{C}$. With these approximations, the original Schr\"odinger Equation \eqref{eq:schrodinger} for electrons can be transformed into a matrix form as
\begin{equation}
    \bm{H} \bm{C}_i = \boldsymbol{\epsilon}_i \bm{S} \bm{C}_i,
    \label{eq:schrodinger_matrix}
\end{equation}
where $\bm{H}$ is the Hamiltonian matrix, $\bm{S}$ is the overlap matrix, and $\boldsymbol{\epsilon}_i$ is the energy for the $i$-th orbital. The Hamiltonian matrix can be decomposed into the sum 
\begin{equation}
    \bm{H} = \bm{H}_{eN} + \bm{H}_{ee} + \bm{H}_{XC},
\end{equation}
which describes electron-nucleus interactions ($\bm{H}_{eN}$), electron-electron interactions ($\bm{H}_{ee}$, including kinetic energy and electron–electron Coulomb repulsion energy), and exchange-correlation energy ($\bm{H}_{XC}$). These matrices take the electron density $\bm{\rho}(\bm{r})$ as an input to evaluate the Hamiltonian matrix. The exchange-correlation energy functional used in this paper was B3LYP \citep{b3lyp_lyp,b3lyp_becke1993}, which is a hybrid functional that includes both the exchange energy from the HF method as well as a correlation potential. Thus, the complexity of using B3LYP in DFT is $O(n^4 T)$, which is the same as HF. We implement the GTO basis set Def2SVP \citep{weigend2005balanced} that incorporates aspects of DFT, namely, an exchange-correlation potential, in order to more accurately capture electron-electron interactions compared to the HF method, which uses a mean field approximation of electron density.  

Equation \eqref{eq:schrodinger_matrix} is satisfied for the final Hamiltonian matrix and its coefficient matrix once self-consistency is achieved using direct inversion in the iterative subspace (DIIS) \citep{pulay1980,pulay1982}. The equation is solved iteratively by building and solving the Hamiltonian and coefficient matrices, constructing an error vector based on a linear combination of energy differences in the previous steps, then diagonalizing and recalculating $\bm{H}$ until the error vector is below a convergence threshold. In our QH9 datasets, we provide the Hamiltonian matrix $\bm{H}$ used to train quantum tensor networks for directly predicting the Hamiltonian matrix.


\subsection{Group Equivariance and Equivariant Matrices}

In many quantum chemistry problems, the molecular property to be predicted (e.g., energy and force) is internally invariant or equivariant to transformations in SE(3) group, including rotations and translations. Formally, for an $n$-atom molecule whose 3D atom coordinates are $\bm{r}_1,...,\bm{r}_n$, any transformation in SE(3) group can be described as changing the 3D atom coordinates to $\bm{Rr}_1+\bm{t},...,\bm{Rr}_n+\bm{t}$. Here, the translation vector $\bm{t}\in\mathbb{R}^3$ is an arbitrary 3D vector, and the rotation matrix $\bm{R}\in\mathbb{R}^{3\times 3}$ satisfies that $\bm{R^TR}=\bm{I}, |\bm{R}|=1$. Let $\bm{f}(\cdot)$ map the 3D atom coordinates to an $(2\ell+1)$-dimensional prediction target vector, we say $\bm{f}$ is order-$\ell$ SE(3)-equivariant if
\begin{equation}
	\label{eqn:se3_equi}
	\bm{f}(\bm{Rr}_1+\bm{t},...,\bm{Rr}_n+\bm{t})=D^\ell(\bm{R})\bm{f}(\bm{r}_1,...,\bm{r}_n)
\end{equation}
holds for any rotation matrix $\bm{R}$ and translation vector $\bm{t}$, where $D^\ell(\bm{R})\in\mathbb{C}^{(2\ell+1)\times(2\ell+1)}$ is the order-$\ell$ Wigner-D matrix of $\bm{R}$ (please refer to Section A.3 of~\citet{brandstetter2022geometric} and Section A.2.1 of~\citet{poulenard2022equivalence} for more background information about the Wigner-D matrix). To accurately predict SE(3)-equivariant properties, an effective approach is to develop neural network models that are designed to maintain the same equivariance relations between inputs and outputs as in Equation~(\ref{eqn:se3_equi}). Recently, many studies have proposed SE(3)-equivariant neural network architectures by using SE(3)-invariant feature encoding~\citep{schutt2018schnet,gasteigerdirectional,gasteiger2021gemnet,liu2021spherical}, tensor product operations~\citep{thomas2018tensor,brandstetter2022geometric,liao2023equiformer}, or atomic cluster expansion framework~\citep{batatia2022mace, drautz2019atomic, dusson2022atomic, kovacs2021linear, musaelian2023learning}.

Different from vector-like molecular properties, the Hamiltonian matrix $\bm{H}$ has a much more complicated SE(3) equivariance pattern that is associated with the intrinsic angular momentum of the atomic orbital pairs. 
In computational quantum chemistry algorithms such as DFT, the Hamiltonian matrix $\bm{H}$ obtained from DFT calculations can be used to represent the interactions between these atomic orbitals, and the block $\bm{H}_{ij}$ in Hamiltonian matrix represents the interactions between the atomic orbitals $i$ in atom $a_i$ with angular quantum number $\ell_i$ and atomic orbitals $j$ in atom $a_j$ with angular quantum number $\ell_j$, and the shape of this block $\bm{H}_{ij}$ is $(2 \ell_i + 1) \times (2 \ell_j + 1)$. Usually, the atomic orbitals are arranged sequentially for the orbitals in the same atom and with the same angular quantum number. For example, $\bm{H}_{ij}$ can be located within the $s_i$-th to $(s_i+2\ell_i)$-th row, and the $s_j$-th to $(s_j+2\ell_j)$-th column of Hamiltonian matrix $\bm{H}$.
Specifically, its SE(3) equivariance can be described as
\begin{equation}
	\label{eqn:se3_equi_mat}
	\bm{H}_{ij}\left(\bm{\rho}(\bm{Rr}+ \bm{t})\right)=D^{\ell_i}(\bm{R})\bm{H}_{ij}\left(\bm{\rho}(\bm{r})\right)D^{\ell_j}(\bm{R}^T),
\end{equation}
where $\bm{\rho}(\bm{r})$ is the electronic density at positioin $\bm{r}$ and Hamiltonian matrix $\bm{H}$ is a function of the electronic density $\bm{\rho}(\bm{r})$ in the DFT algorithm.
In other words, the SE(3) equivariance of different submatrices in $\bm{H}$ has different mathematical forms, which is much more complicated than the SE(3) equivariance of vector-like molecular properties. Hence, it is much more challenging to develop SE(3)-equivariant neural network architectures for the prediction of Hamiltonian matrices. Nowadays, only a few studies~\citep{li2022deep,gong2023general,yu2023efficient, unke2021se} have made initial exploration in this direction. 

\subsection{Datasets for Quantum Chemistry}
To facilitate the usage of machine learning models to predict chemistry properties and accelerate simulations, numerous quantum chemistry datasets have been built to provide extensive and faithful data. 
Here, we introduce several existing datasets that have been constructed for different tasks respectively, including molecular property prediction, catalyst modeling, molecular force field prediction, and molecular Hamiltonian matrix prediction.
For molecular property prediction, the QM7~\citep{blum2009970} dataset was initially constructed using 7,165 molecules from the organic molecule database GDB-13~\citep{blum2009970}, with each selected molecule having no more than 7 heavy atoms. The primary purpose of creating the QM7 dataset is to provide atomization energies as the target molecular property.
Then QM9~\citep{ramakrishnan2014quantum, schutt2018schnet} was built based on GDB-17~\citep{ruddigkeit2012enumeration} to provide 134k stable small organic molecules with no more than 9 heavy atoms in each molecule. Moreover, it provides 13 different important quantum chemistry properties, including HOMO and LUMO energies. 
Based on the molecules from PubQChem~\citep{wang2009pubchem, wang2017pubchem, kim2019pubchem, kim2021pubchem, kim2023pubchem}, PubQChemQC~\citep{nakata2017pubchemqc} provides 3M ground-state molecular structures as well as the HOMO-LUMO gap and excitation energies for 2M molecules.
In addition to the molecular property datasets, OC20~\citep{chanussot2021open} and OC22~\citep{oc22} were developed to provide the data of interactions of catalysts on material surfaces. They provide the geometries of the initial structures to predict the final structures or energies as well as the relaxation trajectories with energy and atomic forces.
For the molecular force field prediction datasets, MD17~\citep{chmiela2017machine} and MD22~\citep{chmiela2023accurate} contain atomic forces for molecular and supramolecular trajectories respectively as valuable datasets to develop machine learning methods. 
The last category is the Hamiltonian matrices datasets. MD17~\citep{schutt2019unifying, gastegger2020deep} provides the Hamiltonian matrices for single molecular dynamic trajectories to study the Hamiltonian matrices for molecules with various geometries. Building upon this dataset, mixed MD17~\citep{yu2023efficient} combines four molecular trajectories in the MD17 to study Hamiltonian matrix prediction tasks with multiple molecules. NablaDFT \citep{khrabrov2022nabladft} has million of Hamiltonian matrices for molecular conformers and provides the MAE on predicted Hamiltonian matrices with model trained with in-distribution data split.
Alongside the increasing interest in Hamiltonian matrix prediction, there is a growing need for datasets that include Hamiltonian matrices with a greater number of molecules and benchmark with comprehensive evaluation to facilitate the subsequent studies.


\section{Datasets, Tasks, Methods, and Metrics}

\subsection{Datasets}

\textbf{Dataset Generation.} For the QH9 dataset, we use open-source software PySCF \citep{sun2018pyscf, sun2020recent} to conduct computational quantum chemistry calculations. In the QH9, there are two sub datasets. The first one is the QH9-stable dataset containing Hamiltonian matrices for 130,831 molecules with their geometries. 
We obtain the molecules and their geometries in QH9-stable from the QM9 version with 130,831, which is widely used in molecular property prediction tasks in the literature~\citep{schutt2018schnet, gasteigerdirectional, liu2021spherical, wang2022comenet}. Note that we cover all the molecules in this QM9 version.
The second one is the QH9-dynamic dataset, it has molecular trajectories for $999$ molecules and each trajectory contains $100$ geometries. 
To obtain the accurate Hamiltonian matrices for this dataset, we set the hyper-parameters of the DFT algorithms to a tight level. Specifically, we set the grid density level to 3 to calculate accurate electronic density, and the SCF convergence condition is set to SCF tolerance of $10^{-13}$ and gradient threshold of $3.16\times 10^{-5}$ to ensure the the final states achieve tight convergence.
For the density functional, we select the B3LYP exchange-correlation functional to conduct DFT calculations, and GTO orbital basis Def2SVP is selected to approximate the electronic wavefunctions. To accelerate and achieve the convergence of SCF algorithm, we use DIIS algorithm with consideration of the 8 previous steps. 
For the QH9-dynamic dataset, molecular dynamics simulations are conducted under the microcanonical ensemble, where the number of particles, volume, and energy remain constant (NVE), and the temperature is set to 300K. For QH9-dynamic-100k, the time step for recording the molecular trajectory is set to $0.12$ fs with $1,000$ total steps and then sampled one step every 10 steps. 
For QH9-dynamic-300k, it conducts MD simulations with time step of $1.2$ fs with $100$ total steps. 
The generated QH9 dataset is available at Zenodo (\url{https://zenodo.org/records/8274793}) and GitHub (\url{https://github.com/divelab/AIRS/tree/main/OpenDFT/QHBench/QH9}).




\textbf{Dataset Statistics.}
The statistical data, including the number of molecules and geometries for QH9-stable and QH9-dynamic, is presented in Table \ref{tab:splits_stats}. These molecules consist of no more than 9 heavy atoms and are composed of four specific heavy atoms: carbon (C), nitrogen (N), oxygen (O), and fluorine (F).
The detailed statistics are shown in Table \ref{tab:statistic_details}.

\subsection{Tasks}
\label{sec:tasks}


\begin{table}[t]
    \centering
    \resizebox{\textwidth}{!}{
    \begin{threeparttable}
    \caption{The statistics of our defined four tasks.}
        \begin{tabular}{l|cccc}
        \toprule 
        Task & \# Total geometries & \# Molecules &  \# Training/validation/testing geometries  \\
        \midrule
        QH9-stable-id         &  $130,831$ & $130,831$ & $ 104,664/13,083/13,084$  \\
        QH9-stable-ood         &  $130,831$ & $130,831$ & $  104,001/17,495/9,335$  \\
        QH9-dynamic-100k-geo \tnote{a}        &   $99,900$ & $999$ & $79,920/9,990/9,990$ \\
        QH9-dynamic-100k-mol         &  $99,900$ & $999$ & $  79,900/9,900/10,100$  \\
        QH9-dynamic-300k-geo  \tnote{b}      & $299,800$ & $2,998$ & $239,840$/$29,980$/$29,980$ \\
        QH9-dynamic-300k-mol        & $299,800$ & $2,998$ & $239,800$/$29,900$/$30,100$ \\
        \bottomrule
        \end{tabular}
     \begin{tablenotes}
        \footnotesize
       \item [a] QH9-dynamic-100k conducts MD simulations with time step of $5$ a.u. ($ \approx 0.120944$ fs) for $1,000$ total steps, and selects one sample every 10 steps.
       \item [b] QH9-dynamic-300k conducts MD simulations with time step of $50$ a.u. ($ \approx 1.20944$ fs) for 100 total steps.
    \end{tablenotes}
    \label{tab:splits_stats}
    \end{threeparttable}
    }
\end{table}

To comprehensively evaluate the quantum Hamiltonian prediction performance, we define the following tasks based on the obtained stable and dynamic geometries in the QH9 dataset.

\textbf{QH9-stable-id.} We first randomly divide the obtained stable geometries in QH9 into three subsets, including $80\%$ for training, $10\%$ for validation, and $10\%$ for testing. This serves as the basic evaluation task for predicting quantum Hamiltonian matrices. 

\textbf{QH9-stable-ood.} We further split the stable geometries in QH9 by molecular size based on the number of constituting atoms. The training set consists of molecules with $3$ to $20$ atoms, maintaining a similar number of training samples as in the QH9-stable-id split. The validation set includes molecules with $21$ to $22$ atoms, while the testing set has molecules with $23$ to $29$ atoms. This task allows for an evaluation of the model's generalization ability under an out-of-distribution training setup.

\textbf{QH9-dynamic-geo.} For this split and the following QH9-dynamic-mol split, they use the molecular dynamics trajectories, while each trajectory includes $100$ geometries. In QH9-dynamic-geo, the split is performed geometry-wise. Specifically, for each molecule, $100$ geometries are randomly divided into $80$ for training, $10$ for validation, and $10$ for testing. Here, the molecules in the test set are visible during training but the geometric structures are different from training structures.

\textbf{QH9-dynamic-mol.} In the QH9-dynamic-mol partitioning, QH9-dynamic-100k-mol and QH9-dynamic-300k-mol datasets utilize $999$ and $2,998$ molecules, respectively, and molecules are divided into training, validation, and testing subsets in a ratio of $0.8/0.1/0.1$. Importantly, different from the above QH9-dynamic-geo setup, all $100$ geometries corresponding to a specific molecule are grouped together and assigned to the same subset. This setup introduces a more challenging task than QH9-dynamic-geo since the geometries in the testing set correspond to different molecules as those in training.

\subsection{Methods}
To predict the quantum Hamiltonian matrix, several quantum tensor networks have been proposed \citep{li2023deep, gong2023general}. SchNOrb \citep{schutt2019unifying} uses pairwise distance and direction as the input geometric information to predict the final Hamiltonian matrix. However, SchNOrb lacks the ability to ensure matrix equivariance and relies on data augmentation techniques to encourage equivariance.
Another network, DeepH \citep{li2022deep}, uses invariant local coordinate systems and a global coordinate system to handle the equivariance challenge. It uses the geometric features within these invariant local coordinate systems to predict invariant Hamiltonian matrix blocks. Next, as a post-processing step, DeepH applies a rotation using Wigner D-Matrix to transform the Hamiltonian matrix blocks from local coordinate systems back to the global coordinate system. Currently, DeepH is applied on predicting Hamiltonian matrices for materials.
PhiSNet \citep{unke2021se} uses an equivariant model architecture that inherently guarantees matrix equivariance.  However, current implementation of PhiSNet is limited to supporting single molecule. This limitation arises from the design of the matrix prediction module in PhiSNet, which is designed to predict matrices for the same molecules with fixed matrix size. 
Therefore, equivariant quantum tensor network QHNet \citep{yu2023efficient} is selected as the main baseline method in the QH9 benchmark currently. QHNet has an extendable expansion module that is built upon intermediate full orbital matrices, enabling its capability to effectively handle different molecules. This flexibility allows QHNet to accommodate various molecules in the QH9 benchmark.

\subsection{Metrics}
\label{sec:metrics}

To evaluate the quality of the predicted Hamiltonian matrix, we adopt several metrics that are used to measure both approximation accuracy and computational efficiency.

\textbf{MAE on Hamiltonian matrix $\mathbf{H}$.} This metric calculates the Mean Absolute Error (MAE) between the predicted Hamiltonian matrix and the ground-truth labels from DFT calculation. Each Hamiltonian matrix consists of diagonal blocks and non-diagonal blocks, representing the interactions within individual atoms and the interactions between pairs of atoms, respectively.  When the atom pair is distant, the values in the Hamiltonian matrix blocks are typically close to zero. Consequently, as the molecules increase in size, the proportion of distant atom pairs also increases, causing the overall mean value of the Hamiltonian matrix to decrease. Hence, in the subsequent experiments, we compare the MAEs of the diagonal and non-diagonal blocks separately as well as the total MAE on the Hamiltonian matrix.

\textbf{MAE on occupied orbital energies $\boldsymbol{\epsilon}$}. Orbital energy, which includes the Highest Occupied Molecular Orbital (HOMO) and Lowest Unoccupied Molecular Orbital (LUMO) energies, is a highly significant chemical property. It can be determined by diagonalizing the Hamiltonian matrix using Equation \ref{eq:schrodinger_matrix}. Hence, this metric can serve as a measure to reflect the quality of the predicted Hamiltonian matrix in accurately deducing the desired property. Specifically, it calculates the MAE on all the occupied molecular orbital energies $\boldsymbol{\epsilon}$ derived from the predicted and the ground-truth Hamiltonian matrix.

\textbf{Cosine similarity of orbital coefficients $\psi$.} Electronic wavefunctions can describe the quantum states of molecular systems and are used to derive a range of chemical properties. In order to measure the similarity between the ground-truth wavefunctions and the predicted wavefunctions, we calculate the cosine similarity of the coefficients for the occupied molecular orbitals $\psi$. The corresponding coefficients $\mathbf{C}$ are derived from the predicted and ground-truth Hamiltonian matrix shown in Equation \ref{eq:schrodinger_matrix}.

\textbf{Acceleration ratio.} Besides the metrics assessing molecular properties, several acceleration ratios, such as achieved ratio and error-level ratio, are proposed to measure the quality of the predicted Hamiltonian matrix in accelerating DFT calculation. 
Specifically, the achieved ratio calculates the ratio of the number of optimization steps between initializing with the predicted Hamiltonian matrix and using initial guess methods like minao and 1e.
When the Hamiltonian matrix is accurately predicted, the Self-Consistent Field (SCF) algorithm approaches convergence, resulting in a substantial reduction in the number of optimization steps.
As a comparison, the optimal ratio calculates the single optimization step for each molecule divided by the total number of steps, serving as an illustrative benchmark of the ideal performance.
Meanwhile, the error-level ratio calculates the ratio over the number of steps required to reach a the same error level as model prediction during the DFT SCF loop compared to the total number of steps in the DFT process.

\section{Experiments} 

\textbf{Setup.} To assess how deep learning approaches perform on the proposed dataset, we conduct experiments on the four designed tasks, as described in Section~\ref{sec:tasks}. To be more specific, we evaluate the performance of QHNet~\citep{yu2023efficient}, a recently proposed SE(3)-equivariant network specifically designed for efficient and accurate quantum Hamiltonian matrix prediction. QHNet is known for its effectiveness and efficiency in handling the task at hand, making it a suitable testing method for our benchmark evaluation. For quantitative evaluation, we use the metrics as introduced in Section~\ref{sec:metrics}. Our implementation is based on PyTorch~\citep{paszke2019pytorch}, PyG~\citep{fey2019fast}, and e3nn~\citep{e3nn}. We train models on either (1) a single 48GB Nvidia GeForce RTX A6000 GPU and Intel Xeon Silver 4214R CPU, or (2) a single Nvidia A100 GPU and  Intel Xeon Gold 6258R CPU.

Following the model setup in QHNet, in all implemented models, we employ five node-wise interaction layers to aggregate messages from neighboring nodes
to update the node irreducible representations.
We train all models with a total training step of either $210,000$ or $260,000$ using a batch size of $32$. To expedite the convergence of model training, following the QHNet setup, we implement a learning rate scheduler. The scheduler gradually increases the learning rate from $0$ to a maximum value of $5 \times 10^{-4}$  over the first $1,000$ warm-up steps. Subsequently, the scheduler linearly reduces the learning rate, ensuring it reaches $1 \times 10^{-7}$  by the final step. 


\begin{table}[t]
    \centering
    
    \caption{(Results updated by using the newest QHNet model.) The overall performance on the testing set on the defined four tasks. The unit for the Hamiltonian $\mathbf{H}$ and eigenenergies $\boldsymbol{\epsilon}$ is Hartree denoted by $E_h$.}
    \label{tab:overall_perf}
    \resizebox{1\columnwidth}{!}{
    \begin{threeparttable}
    \begin{tabular}{lcccccc}
    \toprule 
    \multirow{2}{*}{Dataset} &  \multirow{2}{*}{Model} & \multicolumn{3}{c}{$\mathbf{H}$ $[10^{-6} E_h] \downarrow$}  &  \multirow{2}{*}{$\boldsymbol{\epsilon}$ $[10^{-6} E_h] \downarrow$}  & \multirow{2}{*}{$\psi$ $[10^{-2}] \uparrow$} \\
    & & diagonal & non-diagonal &  all & & \\
    \midrule
    QH9-stable-id         &  QHNet  & $111.21$ & $73.68$ & $76.31$ & $798.51$ & $95.85$\\
     QH9-stable-ood &  QHNet   &  $111.72$ & $69.88$ & $72.11$ & $644.17$ & $93.68$\\
     \midrule
     QH9-dynamic-100k-geo     &  QHNet & $149.62$ & $92.88$ & $96.85$ & $834.47$ & $94.45$\\
     QH9-dynamic-100k-mol     &  QHNet & $416.99$ & $153.68$ & $173.92$ & $9719.58$ & $79.15$ \\
    QH9-dynamic-300k-geo \tnote{a}     &  QHNet & $166.99$ & $95.25$ & $100.19$ & $843.14$ & $94.95$\\
     QH9-dynamic-300k-mol \tnote{a}    &  QHNet & $108.70$ & $261.63$ & $119.66$ & $2178.15$ & $90.72$ \\
    \bottomrule
    \end{tabular}
     \begin{tablenotes}
            \footnotesize
       \item [a] The cost of training on QH9-dynamic-300k is similar compared to QH9-dynamic-100k, while it contains more data and achieves higher performance in molecule-wise split. Therefore, it is recommended to use QH9-dynamic-300k.  
    \end{tablenotes}
    \end{threeparttable}
    }
    \label{tab:performance}
\end{table}

\textbf{Overall performance.} We first evaluate the overall performance of the model on the four defined tasks by demonstrating its accuracy of the predicted Hamiltonian matrices on the testing set. As summarized in Table~\ref{tab:overall_perf}, the employed QHNet models can achieve a reasonably low MAE in predicting the Hamiltonian matrices on all proposed tasks. For reference, QHNet can achieve an MAE of $83.12 \times 10^{-6} E_h$ on the mixed MD17 dataset, which has a similar setup to our QH9-dynamic-geo setup. In addition to MAE on Hamiltonian matrices, the trained models also achieve low errors on the predicted occupied orbital energies and orbital coefficients. This aligns with the prior reported work that QHNet is effective to predict the Hamiltonian matrices for multiple molecules~\citep{yu2023efficient}. Notably, compared to the existing Hamiltonian matrix datasets, such as MD17~\citep{chmiela2017machine} and mixed MD17~\citep{yu2023efficient}, our proposed tasks involve predicting Hamiltonian matrices for significantly more molecules. Overall, we anticipate that the proposed new datasets and corresponding tasks can serve as more challenging and realistic testbeds for future research in Hamiltonian matrix prediction.

\textbf{Investigation on out-of-distribution generalization.} Since we maintain a similar number of training samples for QH9-stable-id and QH9-stable-ood, it is feasible to compare the performance of these two settings to investigate the out-of-distribution challenge in predicting Hamiltonian matrices. It is worth noting that we cannot directly compare the performance on their respective test sets, as reported in Table~\ref{tab:overall_perf}, to demonstrate the out-of-distribution generalizability challenge. This is because the molecules in the QH9-stable-ood test set have a larger number of atoms on average than those in QH9-stable-id. As explained in Section~\ref{sec:metrics}, molecules with larger size typically have more distant atom pairs, thus leading to a lower overall mean value of the Hamiltonian matrix. Hence, numerical results on molecules with different sizes are not directly comparable.

To examine the presence of the out-of-distribution issue in the Hamiltonian prediction task, we adopt an alternative evaluation strategy. To be specific, we assess models that have been trained respectively on the QH9-stable-id and QH9-stable-ood training sets, employing the same set of samples for evaluation in each instance. Specifically, we use the intersecting set of the QH9-stable-id and QH9-stable-ood testing sets as our evaluation dataset. Clearly, the samples contained within this evaluation set are previously unseen during the training phase of either model, thereby maintaining the integrity of the assessment. The evaluation set contains $923$ molecules with $23$ to $29$ atoms. Under this experimental setup, the primary challenge faced by the model trained on the QH9-stable-ood training set stems from the novelty of molecular sizes during the evaluation phase. 
On the other hand, the model trained on the QH9-stable-id training set benefits from having been exposed to such molecular sizes during training. We denote that these two models are trained under out-of-distribution (OOD) and in-distribution (ID) training schema respectively in Table~\ref{tab:ood_perf}. By comparing the performance on the identical evaluation set, it becomes apparent that the model employing the ID training schema outperforms its OOD-trained counterpart, across all metrics including Hamiltonian MAE and predicted orbital energies and coefficients. Such a performance gap demonstrates that the out-of-distribution issue in molecular size is actually a valid concern particularly when extending trained models to molecular sizes not encountered during training.


\begin{table}[t]
    \centering
    \caption{The performance of  in-distribution (ID) training and out-of-distribution (OOD) training on the constructed evaluation set for the OOD investigation.}
    \label{tab:ood_perf}
    \resizebox{1\columnwidth}{!}{
    \begin{tabular}{lcccccc}
    \toprule 
    \multirow{2}{*}{Training schema}  & \multirow{2}{*}{Models} & \multicolumn{3}{c}{$\mathbf{H}$ $[10^{-6} E_h] \downarrow$}  &  \multirow{2}{*}{$\boldsymbol{\epsilon}$ $[10^{-6} E_h] \downarrow$}  & \multirow{2}{*}{$\psi$ $[10^{-2}] \uparrow$} \\
    & & diagonal & non-diagonal &  all & & \\
    \midrule
    ID &    QHNet  & $84.19$ & $56.01$ & $57.53$ & $442.78$ & $95.26$\\ 
     OOD &  QHNet   & $113.05$ & $70.52$ & $72.78$ & $630.49$ & $94.01$\\ 
    \bottomrule
    \end{tabular}
    }
\end{table}

\textbf{Geometry-wise \emph{vs.} molecule-wise generalization.} We further explore geometry-wise and molecule-wise generalizability by analyzing the difficulty differences between the QH9-dynamic-geo and QH9-dynamic-mol tasks. We consider the results in Table~\ref{tab:overall_perf} for these two tasks to be comparable given that both models are trained with a similar number of geometry structures. We note that the model in the QH9-dynamic-geo task demonstrates numerically better test performance than the model in the QH9-dynamic-mol task. This is consistent with our intention when designing the tasks. Specifically, in QH9-dynamic-geo, although the geometric structures in the test set are different, the molecules themselves are not entirely novel to the model due to the exposure during the training phase.
In comparison, the QH9-dynamic-mol task presents a more challenging and demanding scenario. In particular, the test set geometries in QH9-dynamic-mol correspond to entirely different molecules than those seen during training. This task requires the model to generalize from its learned patterns to the unseen molecular structures. 
Notably, due to the increased number of trajectories with more diverse molecules in the QH9-dynamic-300k-mol dataset, the test performance significantly improves compared to the QH9-dynamic-100k-mol under the same training settings and cost.
To summarize,  both tasks serve as valuable testbeds in evaluating the model's generalization ability, and our analysis shows that QH9-dynamic-mol task, which requires extrapolating to entirely new molecular structures, is notably more challenging and demanding.


\begin{table}[t]
    \centering
    \caption{The performance of DFT calculation acceleration. Both models, trained on the QH9-stable-id split and the QH9-stable-ood split respectively, are evaluated on a common set of $50$ randomly chosen molecules from the intersection of their test sets, making their results directly comparable. Similarly, results from models trained on the QH9-dynamic-geo and QH9-dynamic-mol splits can be compared directly. This table demonstrates the possibility of using quantum tensor network as the initialization for quantum chemistry calculations, thereby speeding up the convergence. However, it is not necessary to run this metric for new designed models due to the computational cost of DFT.}
    \label{tab:optimization_ratio}
    \resizebox{\columnwidth}{!}{
    \begin{tabular}{lccc|lccc}
    \toprule 
    Training Dataset & DFT initialization & Metric & Ratio & Training Dataset & DFT initialization & Metric & Ratio      \\
    \midrule
    \multirow{6}{*}{QH9-stable-id} & \multirow{3}{*}{1e} & Optimal ratio & $0.057 \scriptstyle{{\light{\pm0.004}}}$ & \multirow{6}{*}{QH9-stable-ood} & \multirow{3}{*}{1e} & Optimal ratio & $0.057 \scriptstyle{{\light{\pm0.004}}}$ \\
    & & Achieved ratio $\downarrow$ & $\textbf{0.395} \scriptstyle{{\light{\pm0.030}}}$ & & & Achieved ratio $\downarrow$ & $0.400 \scriptstyle{{\light{\pm0.030}}}$   \\
    & & Error-level ratio $\uparrow$ & $\textbf{0.635} \scriptstyle{{\light{\pm0.039}}}$ & & & Error-level ratio $\uparrow$ & $0.620 \scriptstyle{{\light{\pm0.037}}}$  \\
     & \multirow{3}{*}{minao} & Optimal ratio & $0.102 \scriptstyle{{\light{\pm0.005}}}$& & \multirow{3}{*}{minao} & Optimal ratio & $0.102 \scriptstyle{{\light{\pm0.005}}}$ \\
    & & Achieved ratio $\downarrow$ & $\textbf{0.706} \scriptstyle{{\light{\pm0.031}}}$ & &  & Achieved ratio $\downarrow$ & $0.715 \scriptstyle{{\light{\pm0.033}}}$ \\
    & & Error-level ratio $\uparrow$ & $\textbf{0.408} \scriptstyle{{\light{\pm0.025}}}$ & & & Error-level ratio $\uparrow$ & $0.406 \scriptstyle{{\light{\pm0.021}}}$ \\
    \midrule
    \multirow{6}{*}{QH9-dynamic-100k-geo} & \multirow{3}{*}{1e} & Optimal ratio & $0.056 \scriptstyle{{\light{\pm0.006}}}$ & \multirow{6}{*}{QH9-dynamic-100k-mol} & \multirow{3}{*}{1e} & Optimal ratio & $0.056 \scriptstyle{{\light{\pm0.006}}}$  \\
    & & Achieved ratio $\downarrow$ &  $\textbf{0.392} \scriptstyle{{\light{\pm0.036}}}$ & & & Achieved ratio $\downarrow$ & $0.512 \scriptstyle{{\light{\pm0.138}}}$ \\
    & & Error-level ratio $\uparrow$ &  $\textbf{0.648} \scriptstyle{{\light{\pm0.041}}}$ & & & Error-level ratio $\uparrow$ & $0.622 \scriptstyle{{\light{\pm0.048}}}$ \\
     & \multirow{3}{*}{minao} & Optimal ratio &  $0.098 \scriptstyle{{\light{\pm0.008}}}$ & & \multirow{3}{*}{minao} & Optimal ratio & $0.098 \scriptstyle{{\light{\pm0.008}}}$ \\
    & & Achieved ratio $\downarrow$ & $\textbf{ 0.679} \scriptstyle{{\light{\pm0.041}}}$ & &  & Achieved ratio $\downarrow$ & $ 0.882 \scriptstyle{{\light{\pm0.217}}}$ \\
    & & Error-level ratio $\uparrow$ & $\textbf{0.443} \scriptstyle{{\light{\pm0.044}}}$ & & & Error-level ratio $\uparrow$ &  $0.406 \scriptstyle{{\light{\pm0.066}}}$ \\
    \midrule
    \multirow{6}{*}{QH9-dynamic-300k-geo} & \multirow{3}{*}{1e} & Optimal ratio & $0.057 \scriptstyle{{\light{\pm0.007}}}$ & \multirow{6}{*}{QH9-dynamic-300k-mol} & \multirow{3}{*}{1e} & Optimal ratio & $0.057 \scriptstyle{{\light{\pm0.007}}}$  \\
    & & Achieved ratio $\downarrow$ &  $\textbf{0.392} \scriptstyle{{\light{\pm0.036}}}$ & & & Achieved ratio $\downarrow$ & $0.512 \scriptstyle{{\light{\pm0.138}}}$ \\
    & & Error-level ratio $\uparrow$ &  $\textbf{0.648} \scriptstyle{{\light{\pm0.041}}}$ & & & Error-level ratio $\uparrow$ & $0.622 \scriptstyle{{\light{\pm0.048}}}$ \\
     & \multirow{3}{*}{minao} & Optimal ratio &  $0.098 \scriptstyle{{\light{\pm0.008}}}$ & & \multirow{3}{*}{minao} & Optimal ratio & $0.098 \scriptstyle{{\light{\pm0.008}}}$ \\
    & & Achieved ratio $\downarrow$ & $\textbf{ 0.679} \scriptstyle{{\light{\pm0.041}}}$ & &  & Achieved ratio $\downarrow$ & $ 0.882 \scriptstyle{{\light{\pm0.217}}}$ \\
    & & Error-level ratio $\uparrow$ & $\textbf{0.443} \scriptstyle{{\light{\pm0.044}}}$ & & & Error-level ratio $\uparrow$ &  $0.406 \scriptstyle{{\light{\pm0.066}}}$ \\
    
    \bottomrule
    \end{tabular}
    }
\end{table}

\textbf{Accelerating the DFT calculation.} We further measure the quality of the predicted Hamiltonian matrix by evaluating its ability in accelerating the DFT calculation. As introduced in Section~\ref{sec:metrics}, we compute the ratio of optimization steps required when initializing with the predicted Hamiltonian matrix as compared to classic initial guess methods such as minao and 1e. Note that minao diagonalizes the Fock matrix obtained from a minimal basis to get the guess orbitals, and 1e is one-electron guess which diagonalizes the core Hamiltonian to obtain the guess orbitals. In this experiment, following our data collection process, we use PySCF~\citep{sun2018pyscf} to perform the DFT calculation with using B3LYP exchange-correlation functional and def2SVP basis set. We select DIIS as the SCF algorithm for the DFT calculation and set a grid density level of $3$ to ensure an accurate DFT calculation. For each dataset, we compute the average optimization step ratio for $50$ randomly selected molecules. 
As shown in Table~\ref{tab:optimization_ratio}, we provide several metrics to reflect the optimization ratio. The optimal ratio is the ratio of a single step to the DFT optimization steps. The achieve ratio is the number of optimization steps when initialized by model prediction to optimization steps by DFT initialization. The error-level ratio is the number of optimization steps that achieve similar MAE error with model prediction to the total DFT optimization steps. We can observe that when initializing from the predicted Hamiltonian matrices given by QHNet, it requires fewer optimization steps to reach the converged Hamiltonian matrix, which indicates that the predicted Hamiltonian matrix is close to the convergence condition. This set of experimental results demonstrates that machine learning approaches are helpful in accelerating the DFT calculation.

     

\section{Conclusion}

We are interested in accelerating computation of quantum Hamiltonian matrices, which fundamentally determine the quantum
states of physical systems and chemical properties. While various invariant and equivariant deep learning methods have been developed recently, current quantum Hamiltonian datasets consist of Hamiltonian matrices of molecular dynamic trajectories for only a single and four molecules, respectively. To significantly expand the size and variety of such datasets, we generate a much larger dataset based on the QM9 molecules. Our dataset provides precise Hamiltonian matrices for 130,831 stable molecular geometries and $999$ molecular dynamics trajectories with sampled $100$ geometries in each trajectory. Extensive and carefully designed experiments are conducted to demonstrate the quality of our generated data. 


\section*{Acknowledgements}
This work was supported in part by National Science Foundation grants IIS-2006861, CCF-1553281, DMR-2119103, DMR-2103842, CMMI-2226908, and IIS-2212419 and by the donors of ACS Petroleum Research Fund under Grant 65502-ND10. Portions of this research were conducted with the advanced computing resources provided by Texas A\&M High Performance Research Computing.

\newpage
\bibliographystyle{plainnat}
\bibliography{reference,dive}

\begin{thebibliography}{67}
\providecommand{\natexlab}[1]{#1}
\providecommand{\url}[1]{\texttt{#1}}
\expandafter\ifx\csname urlstyle\endcsname\relax
  \providecommand{\doi}[1]{doi: #1}\else
  \providecommand{\doi}{doi: \begingroup \urlstyle{rm}\Url}\fi

\bibitem[Anderson et~al.(2019)Anderson, Hy, and Kondor]{anderson2019cormorant}
Brandon Anderson, Truong~Son Hy, and Risi Kondor.
\newblock Cormorant: Covariant molecular neural networks.
\newblock \emph{Advances in neural information processing systems}, 32, 2019.

\bibitem[Bai et~al.(2022)Bai, Chu, Tsai, Wilson, Qian, Yan, and Ling]{Bai2022graph}
Hexin Bai, Peng Chu, Jeng-Yuan Tsai, Nathan Wilson, Xiaofeng Qian, Qimin Yan, and Haibin Ling.
\newblock {Graph neural network for Hamiltonian-based material property prediction}.
\newblock \emph{Neural Computing and Applications}, 34:\penalty0 4625--4632, 2022.

\bibitem[Batatia et~al.(2022)Batatia, Kovacs, Simm, Ortner, and Csanyi]{batatia2022mace}
Ilyes Batatia, David~Peter Kovacs, Gregor N.~C. Simm, Christoph Ortner, and Gabor Csanyi.
\newblock {MACE}: Higher order equivariant message passing neural networks for fast and accurate force fields.
\newblock In \emph{Advances in Neural Information Processing Systems}, 2022.
\newblock URL \url{https://openreview.net/forum?id=YPpSngE-ZU}.

\bibitem[Batzner et~al.(2022)Batzner, Musaelian, Sun, Geiger, Mailoa, Kornbluth, Molinari, Smidt, and Kozinsky]{batzner20223}
Simon Batzner, Albert Musaelian, Lixin Sun, Mario Geiger, Jonathan~P Mailoa, Mordechai Kornbluth, Nicola Molinari, Tess~E Smidt, and Boris Kozinsky.
\newblock E(3)-equivariant graph neural networks for data-efficient and accurate interatomic potentials.
\newblock \emph{Nature Communications}, 13\penalty0 (1):\penalty0 2453, 2022.

\bibitem[Becke(1993)]{b3lyp_becke1993}
Axel~D. Becke.
\newblock {Density‐functional thermochemistry. III. The role of exact exchange}.
\newblock \emph{The Journal of Chemical Physics}, 98\penalty0 (7):\penalty0 5648--5652, 04 1993.
\newblock ISSN 0021-9606.
\newblock \doi{10.1063/1.464913}.
\newblock URL \url{https://doi.org/10.1063/1.464913}.

\bibitem[Blum and Reymond(2009)]{blum2009970}
Lorenz~C Blum and Jean-Louis Reymond.
\newblock 970 million druglike small molecules for virtual screening in the chemical universe database gdb-13.
\newblock \emph{Journal of the American Chemical Society}, 131\penalty0 (25):\penalty0 8732--8733, 2009.

\bibitem[Brandstetter et~al.(2022)Brandstetter, Hesselink, van~der Pol, Bekkers, and Welling]{brandstetter2022geometric}
Johannes Brandstetter, Rob Hesselink, Elise van~der Pol, Erik~J Bekkers, and Max Welling.
\newblock Geometric and physical quantities improve e(3) equivariant message passing.
\newblock In \emph{International Conference on Learning Representations}, 2022.

\bibitem[Chanussot et~al.(2021)Chanussot, Das, Goyal, Lavril, Shuaibi, Riviere, Tran, Heras-Domingo, Ho, Hu, et~al.]{chanussot2021open}
Lowik Chanussot, Abhishek Das, Siddharth Goyal, Thibaut Lavril, Muhammed Shuaibi, Morgane Riviere, Kevin Tran, Javier Heras-Domingo, Caleb Ho, Weihua Hu, et~al.
\newblock Open catalyst 2020 (oc20) dataset and community challenges.
\newblock \emph{ACS Catalysis}, 11\penalty0 (10):\penalty0 6059--6072, 2021.

\bibitem[Chmiela et~al.(2017)Chmiela, Tkatchenko, Sauceda, Poltavsky, Sch{\"u}tt, and M{\"u}ller]{chmiela2017machine}
Stefan Chmiela, Alexandre Tkatchenko, Huziel~E Sauceda, Igor Poltavsky, Kristof~T Sch{\"u}tt, and Klaus-Robert M{\"u}ller.
\newblock Machine learning of accurate energy-conserving molecular force fields.
\newblock \emph{Science Advances}, 3\penalty0 (5):\penalty0 e1603015, 2017.

\bibitem[Chmiela et~al.(2023)Chmiela, Vassilev-Galindo, Unke, Kabylda, Sauceda, Tkatchenko, and M{\"u}ller]{chmiela2023accurate}
Stefan Chmiela, Valentin Vassilev-Galindo, Oliver~T Unke, Adil Kabylda, Huziel~E Sauceda, Alexandre Tkatchenko, and Klaus-Robert M{\"u}ller.
\newblock Accurate global machine learning force fields for molecules with hundreds of atoms.
\newblock \emph{Science Advances}, 9\penalty0 (2):\penalty0 eadf0873, 2023.

\bibitem[Drautz(2019)]{drautz2019atomic}
Ralf Drautz.
\newblock Atomic cluster expansion for accurate and transferable interatomic potentials.
\newblock \emph{Physical Review B}, 99\penalty0 (1):\penalty0 014104, 2019.

\bibitem[Dusson et~al.(2022)Dusson, Bachmayr, Cs{\'a}nyi, Drautz, Etter, van~der Oord, and Ortner]{dusson2022atomic}
Genevieve Dusson, Markus Bachmayr, G{\'a}bor Cs{\'a}nyi, Ralf Drautz, Simon Etter, Cas van~der Oord, and Christoph Ortner.
\newblock Atomic cluster expansion: Completeness, efficiency and stability.
\newblock \emph{Journal of Computational Physics}, 454:\penalty0 110946, 2022.

\bibitem[Fey and Lenssen(2019)]{fey2019fast}
Matthias Fey and Jan~Eric Lenssen.
\newblock Fast graph representation learning with pytorch geometric.
\newblock \emph{arXiv Preprint, arXiv:1903.02428}, 2019.

\bibitem[Fuchs et~al.(2020)Fuchs, Worrall, Fischer, and Welling]{fuchs2020se}
Fabian Fuchs, Daniel Worrall, Volker Fischer, and Max Welling.
\newblock {SE(3)}-transformers: {3D} roto-translation equivariant attention networks.
\newblock \emph{Advances in Neural Information Processing Systems}, 33:\penalty0 1970--1981, 2020.

\bibitem[Gastegger et~al.(2020)Gastegger, McSloy, Luya, Sch{\"u}tt, and Maurer]{gastegger2020deep}
Michael Gastegger, Adam McSloy, M~Luya, Kristof~T Sch{\"u}tt, and Reinhard~J Maurer.
\newblock A deep neural network for molecular wave functions in quasi-atomic minimal basis representation.
\newblock \emph{The Journal of Chemical Physics}, 153\penalty0 (4):\penalty0 044123, 2020.

\bibitem[Gasteiger et~al.(2020)Gasteiger, Gro{\ss}, and G{\"u}nnemann]{gasteigerdirectional}
Johannes Gasteiger, Janek Gro{\ss}, and Stephan G{\"u}nnemann.
\newblock Directional message passing for molecular graphs.
\newblock In \emph{International Conference on Learning Representations}, 2020.

\bibitem[Gasteiger et~al.(2021)Gasteiger, Becker, and G{\"u}nnemann]{gasteiger2021gemnet}
Johannes Gasteiger, Florian Becker, and Stephan G{\"u}nnemann.
\newblock {GemNet}: Universal directional graph neural networks for molecules.
\newblock \emph{Advances in Neural Information Processing Systems}, 34:\penalty0 6790--6802, 2021.

\bibitem[Geiger et~al.(2022)Geiger, Smidt, M., Miller, Boomsma, Dice, Lapchevskyi, Weiler, Tyszkiewicz, Batzner, Madisetti, Uhrin, Frellsen, Jung, Sanborn, Wen, Rackers, Rød, and Bailey]{e3nn}
Mario Geiger, Tess Smidt, Alby M., Benjamin~Kurt Miller, Wouter Boomsma, Bradley Dice, Kostiantyn Lapchevskyi, Maurice Weiler, Michał Tyszkiewicz, Simon Batzner, Dylan Madisetti, Martin Uhrin, Jes Frellsen, Nuri Jung, Sophia Sanborn, Mingjian Wen, Josh Rackers, Marcel Rød, and Michael Bailey.
\newblock Euclidean neural networks: e3nn, April 2022.
\newblock URL \url{https://doi.org/10.5281/zenodo.6459381}.

\bibitem[Gong et~al.(2023)Gong, Li, Zou, Xu, Duan, and Xu]{gong2023general}
Xiaoxun Gong, He~Li, Nianlong Zou, Runzhang Xu, Wenhui Duan, and Yong Xu.
\newblock {General framework for E(3)-equivariant neural network representation of density functional theory Hamiltonian}.
\newblock \emph{Nature Communications}, 14\penalty0 (1):\penalty0 2848, 2023.

\bibitem[Hohenberg and Kohn(1964)]{hohenkohn}
P.~Hohenberg and W.~Kohn.
\newblock Inhomogeneous electron gas.
\newblock \emph{Phys. Rev.}, 136:\penalty0 B864--B871, Nov 1964.
\newblock \doi{10.1103/PhysRev.136.B864}.
\newblock URL \url{https://link.aps.org/doi/10.1103/PhysRev.136.B864}.

\bibitem[Huang et~al.(2023)Huang, von Rudorff, and von Lilienfeld]{huang2023central}
Bing Huang, Guido~Falk von Rudorff, and O~Anatole von Lilienfeld.
\newblock The central role of density functional theory in the ai age.
\newblock \emph{Science}, 381\penalty0 (6654):\penalty0 170--175, 2023.

\bibitem[Khrabrov et~al.(2022)Khrabrov, Shenbin, Ryabov, Tsypin, Telepov, Alekseev, Grishin, Strashnov, Zhilyaev, Nikolenko, et~al.]{khrabrov2022nabladft}
Kuzma Khrabrov, Ilya Shenbin, Alexander Ryabov, Artem Tsypin, Alexander Telepov, Anton Alekseev, Alexander Grishin, Pavel Strashnov, Petr Zhilyaev, Sergey Nikolenko, et~al.
\newblock nabladft: Large-scale conformational energy and hamiltonian prediction benchmark and dataset.
\newblock \emph{Physical Chemistry Chemical Physics}, 24\penalty0 (42):\penalty0 25853--25863, 2022.

\bibitem[Kim et~al.(2019)Kim, Chen, Cheng, Gindulyte, He, He, Li, Shoemaker, Thiessen, Yu, et~al.]{kim2019pubchem}
Sunghwan Kim, Jie Chen, Tiejun Cheng, Asta Gindulyte, Jia He, Siqian He, Qingliang Li, Benjamin~A Shoemaker, Paul~A Thiessen, Bo~Yu, et~al.
\newblock {PubChem} 2019 update: improved access to chemical data.
\newblock \emph{Nucleic Acids Research}, 47\penalty0 (D1):\penalty0 D1102--D1109, 2019.

\bibitem[Kim et~al.(2021)Kim, Chen, Cheng, Gindulyte, He, He, Li, Shoemaker, Thiessen, Yu, et~al.]{kim2021pubchem}
Sunghwan Kim, Jie Chen, Tiejun Cheng, Asta Gindulyte, Jia He, Siqian He, Qingliang Li, Benjamin~A Shoemaker, Paul~A Thiessen, Bo~Yu, et~al.
\newblock {PubChem} in 2021: new data content and improved web interfaces.
\newblock \emph{Nucleic Acids Research}, 49\penalty0 (D1):\penalty0 D1388--D1395, 2021.

\bibitem[Kim et~al.(2023)Kim, Chen, Cheng, Gindulyte, He, He, Li, Shoemaker, Thiessen, Yu, et~al.]{kim2023pubchem}
Sunghwan Kim, Jie Chen, Tiejun Cheng, Asta Gindulyte, Jia He, Siqian He, Qingliang Li, Benjamin~A Shoemaker, Paul~A Thiessen, Bo~Yu, et~al.
\newblock {PubChem} 2023 update.
\newblock \emph{Nucleic Acids Research}, 51\penalty0 (D1):\penalty0 D1373--D1380, 2023.

\bibitem[Kirkpatrick et~al.(2021)Kirkpatrick, McMorrow, Turban, Gaunt, Spencer, Matthews, Obika, Thiry, Fortunato, Pfau, et~al.]{kirkpatrick2021pushing}
James Kirkpatrick, Brendan McMorrow, David~HP Turban, Alexander~L Gaunt, James~S Spencer, Alexander~GDG Matthews, Annette Obika, Louis Thiry, Meire Fortunato, David Pfau, et~al.
\newblock Pushing the frontiers of density functionals by solving the fractional electron problem.
\newblock \emph{Science}, 374\penalty0 (6573):\penalty0 1385--1389, 2021.

\bibitem[Kohn and Sham(1965)]{kohnsham}
W.~Kohn and L.~J. Sham.
\newblock Self-consistent equations including exchange and correlation effects.
\newblock \emph{Phys. Rev.}, 140:\penalty0 A1133--A1138, Nov 1965.
\newblock \doi{10.1103/PhysRev.140.A1133}.
\newblock URL \url{https://link.aps.org/doi/10.1103/PhysRev.140.A1133}.

\bibitem[Kov{\'a}cs et~al.(2021)Kov{\'a}cs, Oord, Kucera, Allen, Cole, Ortner, and Cs{\'a}nyi]{kovacs2021linear}
D{\'a}vid~P{\'e}ter Kov{\'a}cs, Cas van~der Oord, Jiri Kucera, Alice~EA Allen, Daniel~J Cole, Christoph Ortner, and G{\'a}bor Cs{\'a}nyi.
\newblock {Linear Atomic Cluster Expansion Force Fields for Organic Molecules: Beyond RMSE}.
\newblock \emph{Journal of Chemical Theory and Computation}, 17\penalty0 (12):\penalty0 7696--7711, 2021.

\bibitem[Lee et~al.(1988)Lee, Yang, and Parr]{b3lyp_lyp}
Chengteh Lee, Weitao Yang, and Robert~G. Parr.
\newblock Development of the colle-salvetti correlation-energy formula into a functional of the electron density.
\newblock \emph{Phys. Rev. B}, 37:\penalty0 785--789, Jan 1988.
\newblock \doi{10.1103/PhysRevB.37.785}.
\newblock URL \url{https://link.aps.org/doi/10.1103/PhysRevB.37.785}.

\bibitem[Li et~al.(2022)Li, Wang, Zou, Ye, Xu, Gong, Duan, and Xu]{li2022deep}
He~Li, Zun Wang, Nianlong Zou, Meng Ye, Runzhang Xu, Xiaoxun Gong, Wenhui Duan, and Yong Xu.
\newblock Deep-learning density functional theory hamiltonian for efficient ab initio electronic-structure calculation.
\newblock \emph{Nature Computational Science}, 2\penalty0 (6):\penalty0 367--377, 2022.

\bibitem[Li et~al.(2023)Li, Tang, Gong, Zou, Duan, and Xu]{li2023deep}
He~Li, Zechen Tang, Xiaoxun Gong, Nianlong Zou, Wenhui Duan, and Yong Xu.
\newblock Deep-learning electronic-structure calculation of magnetic superstructures.
\newblock \emph{Nature Computational Science}, 3\penalty0 (4):\penalty0 321--327, 2023.

\bibitem[Liao and Smidt(2023)]{liao2023equiformer}
Yi-Lun Liao and Tess Smidt.
\newblock Equiformer: Equivariant graph attention transformer for 3d atomistic graphs.
\newblock In \emph{The Eleventh International Conference on Learning Representations}, 2023.
\newblock URL \url{https://openreview.net/forum?id=KwmPfARgOTD}.

\bibitem[Liao et~al.(2023)Liao, Wood, Das, and Smidt]{liao2023equiformerv2}
Yi-Lun Liao, Brandon Wood, Abhishek Das, and Tess Smidt.
\newblock Equiformerv2: Improved equivariant transformer for scaling to higher-degree representations.
\newblock \emph{arXiv preprint arXiv:2306.12059}, 2023.

\bibitem[Lin et~al.(2023)Lin, Yan, Luo, Liu, Qian, and Ji]{lin2023efficient}
Yuchao Lin, Keqiang Yan, Youzhi Luo, Yi~Liu, Xiaoning Qian, and Shuiwang Ji.
\newblock Efficient approximations of complete interatomic potentials for crystal property prediction.
\newblock In \emph{Proceedings of the 40th International Conference on Machine Learning}, 2023.

\bibitem[Liu et~al.(2021)Liu, Luo, Wang, Xie, Yuan, Gui, Yu, Xu, Zhang, Liu, Yan, Liu, Fu, Oztekin, Zhang, and Ji]{liu2021dig}
Meng Liu, Youzhi Luo, Limei Wang, Yaochen Xie, Hao Yuan, Shurui Gui, Haiyang Yu, Zhao Xu, Jingtun Zhang, Yi~Liu, Keqiang Yan, Haoran Liu, Cong Fu, Bora~M Oztekin, Xuan Zhang, and Shuiwang Ji.
\newblock {DIG}: A turnkey library for diving into graph deep learning research.
\newblock \emph{Journal of Machine Learning Research}, 22\penalty0 (240):\penalty0 1--9, 2021.
\newblock URL \url{http://jmlr.org/papers/v22/21-0343.html}.

\bibitem[Liu et~al.(2022)Liu, Wang, Liu, Lin, Zhang, Oztekin, and Ji]{liu2021spherical}
Yi~Liu, Limei Wang, Meng Liu, Yuchao Lin, Xuan Zhang, Bora Oztekin, and Shuiwang Ji.
\newblock Spherical message passing for {3D} molecular graphs.
\newblock In \emph{International Conference on Learning Representations}, 2022.

\bibitem[Marzari and Vanderbilt(1997)]{Marzari1997mlwf}
Nicola Marzari and David Vanderbilt.
\newblock {Maximally localized generalized Wannier functions for composite energy bands}.
\newblock \emph{Phys. Rev. B}, 56:\penalty0 12847--12865, 1997.

\bibitem[Marzari et~al.(2012)Marzari, Mostofi, Yates, Souza, and Vanderbilt]{Marzari2012rmp}
Nicola Marzari, Arash~A. Mostofi, Jonathan~R. Yates, Ivo Souza, and David Vanderbilt.
\newblock {Maximally localized Wannier functions: Theory and applications}.
\newblock \emph{Rev. Mod. Phys.}, 84:\penalty0 1419--1475, 2012.

\bibitem[Musaelian et~al.(2023)Musaelian, Batzner, Johansson, Sun, Owen, Kornbluth, and Kozinsky]{musaelian2023learning}
Albert Musaelian, Simon Batzner, Anders Johansson, Lixin Sun, Cameron~J Owen, Mordechai Kornbluth, and Boris Kozinsky.
\newblock Learning local equivariant representations for large-scale atomistic dynamics.
\newblock \emph{Nature Communications}, 14\penalty0 (1):\penalty0 579, 2023.

\bibitem[Nakata and Shimazaki(2017)]{nakata2017pubchemqc}
Maho Nakata and Tomomi Shimazaki.
\newblock {{PubChemQC} Project: A Large-Scale First-Principles Electronic Structure Database for Data-Driven Chemistry}.
\newblock \emph{Journal of Chemical Information and Modeling}, 57\penalty0 (6):\penalty0 1300--1308, 2017.

\bibitem[Paszke et~al.(2019)Paszke, Gross, Massa, Lerer, Bradbury, Chanan, Killeen, Lin, Gimelshein, Antiga, et~al.]{paszke2019pytorch}
Adam Paszke, Sam Gross, Francisco Massa, Adam Lerer, James Bradbury, Gregory Chanan, Trevor Killeen, Zeming Lin, Natalia Gimelshein, Luca Antiga, et~al.
\newblock {PyTorch}: An imperative style, high-performance deep learning library.
\newblock \emph{Advances in Neural Information Processing Systems}, 32, 2019.

\bibitem[Poulenard et~al.(2022)Poulenard, Ovsjanikov, and Guibas]{poulenard2022equivalence}
Adrien Poulenard, Maks Ovsjanikov, and Leonidas~J Guibas.
\newblock Equivalence between se (3) equivariant networks via steerable kernels and group convolution.
\newblock \emph{arXiv preprint arXiv:2211.15903}, 2022.

\bibitem[Pulay(1980)]{pulay1980}
Péter Pulay.
\newblock Convergence acceleration of iterative sequences. the case of scf iteration.
\newblock \emph{Chemical Physics Letters}, 73\penalty0 (2):\penalty0 393--398, 1980.
\newblock ISSN 0009-2614.
\newblock \doi{https://doi.org/10.1016/0009-2614(80)80396-4}.
\newblock URL \url{https://www.sciencedirect.com/science/article/pii/0009261480803964}.

\bibitem[Pulay(1982)]{pulay1982}
Péter Pulay.
\newblock Improved scf convergence acceleration.
\newblock \emph{Journal of Computational Chemistry}, 3\penalty0 (4):\penalty0 556--560, 1982.
\newblock \doi{https://doi.org/10.1002/jcc.540030413}.
\newblock URL \url{https://onlinelibrary.wiley.com/doi/abs/10.1002/jcc.540030413}.

\bibitem[Qian et~al.(2010)Qian, Li, and Yip]{qian2010qotransport}
Xiaofeng Qian, Ju~Li, and Sidney Yip.
\newblock {Calculating phase-coherent quantum transport in nanoelectronics with \textit{ab initio} quasiatomic orbital basis set}.
\newblock \emph{Phys. Rev. B}, 82:\penalty0 195442, 2010.

\bibitem[Ramakrishnan et~al.(2014)Ramakrishnan, Dral, Rupp, and Von~Lilienfeld]{ramakrishnan2014quantum}
Raghunathan Ramakrishnan, Pavlo~O Dral, Matthias Rupp, and O~Anatole Von~Lilienfeld.
\newblock Quantum chemistry structures and properties of 134 kilo molecules.
\newblock \emph{Scientific Data}, 1\penalty0 (1):\penalty0 1--7, 2014.

\bibitem[Ruddigkeit et~al.(2012)Ruddigkeit, Van~Deursen, Blum, and Reymond]{ruddigkeit2012enumeration}
Lars Ruddigkeit, Ruud Van~Deursen, Lorenz~C Blum, and Jean-Louis Reymond.
\newblock {Enumeration of 166 Billion Organic Small Molecules in the Chemical Universe Database {GDB-17}}.
\newblock \emph{Journal of Chemical Information and Modeling}, 52\penalty0 (11):\penalty0 2864--2875, 2012.

\bibitem[Satorras et~al.(2021)Satorras, Hoogeboom, and Welling]{satorras2021n}
V{\i}ctor~Garcia Satorras, Emiel Hoogeboom, and Max Welling.
\newblock E(n) equivariant graph neural networks.
\newblock In \emph{International Conference on Machine Learning}, pages 9323--9332. PMLR, 2021.

\bibitem[Schr{\"o}dinger(1926)]{schrodinger1926undulatory}
Erwin Schr{\"o}dinger.
\newblock An undulatory theory of the mechanics of atoms and molecules.
\newblock \emph{Physical Review}, 28\penalty0 (6):\penalty0 1049, 1926.

\bibitem[Sch{\"u}tt et~al.(2021)Sch{\"u}tt, Unke, and Gastegger]{schutt2021equivariant}
Kristof Sch{\"u}tt, Oliver Unke, and Michael Gastegger.
\newblock Equivariant message passing for the prediction of tensorial properties and molecular spectra.
\newblock In \emph{International Conference on Machine Learning}, pages 9377--9388. PMLR, 2021.

\bibitem[Sch{\"u}tt et~al.(2018)Sch{\"u}tt, Sauceda, Kindermans, Tkatchenko, and M{\"u}ller]{schutt2018schnet}
Kristof~T Sch{\"u}tt, Huziel~E Sauceda, P-J Kindermans, Alexandre Tkatchenko, and K-R M{\"u}ller.
\newblock {SchNet}--a deep learning architecture for molecules and materials.
\newblock \emph{The Journal of Chemical Physics}, 148\penalty0 (24):\penalty0 241722, 2018.

\bibitem[Sch{\"u}tt et~al.(2019)Sch{\"u}tt, Gastegger, Tkatchenko, M{\"u}ller, and Maurer]{schutt2019unifying}
Kristof~T Sch{\"u}tt, Michael Gastegger, Alexandre Tkatchenko, K-R M{\"u}ller, and Reinhard~J Maurer.
\newblock Unifying machine learning and quantum chemistry with a deep neural network for molecular wavefunctions.
\newblock \emph{Nature Communications}, 10\penalty0 (1):\penalty0 5024, 2019.

\bibitem[Souza et~al.(2001)Souza, Marzari, and Vanderbilt]{Souza2001mlwf}
Ivo Souza, Nicola Marzari, and David Vanderbilt.
\newblock {Maximally localized Wannier functions for entangled energy bands}.
\newblock \emph{Phys. Rev. B}, 65:\penalty0 035109, 2001.

\bibitem[Sun et~al.(2018)Sun, Berkelbach, Blunt, Booth, Guo, Li, Liu, McClain, Sayfutyarova, Sharma, et~al.]{sun2018pyscf}
Qiming Sun, Timothy~C Berkelbach, Nick~S Blunt, George~H Booth, Sheng Guo, Zhendong Li, Junzi Liu, James~D McClain, Elvira~R Sayfutyarova, Sandeep Sharma, et~al.
\newblock {PySCF: the Python-based simulations of chemistry framework}.
\newblock \emph{Wiley Interdisciplinary Reviews: Computational Molecular Science}, 8\penalty0 (1):\penalty0 e1340, 2018.
\newblock \doi{https://doi.org/10.1002/wcms.1340}.

\bibitem[Sun et~al.(2020)Sun, Zhang, Banerjee, Bao, Barbry, Blunt, Bogdanov, Booth, Chen, Cui, et~al.]{sun2020recent}
Qiming Sun, Xing Zhang, Samragni Banerjee, Peng Bao, Marc Barbry, Nick~S Blunt, Nikolay~A Bogdanov, George~H Booth, Jia Chen, Zhi-Hao Cui, et~al.
\newblock {Recent developments in the PySCF program package}.
\newblock \emph{The Journal of Chemical Physics}, 153\penalty0 (2), 2020.
\newblock \doi{https://doi.org/10.1063/5.0006074}.

\bibitem[Szabo and Ostlund(2012)]{szabo2012modern}
Attila Szabo and Neil~S Ostlund.
\newblock \emph{{Modern Quantum Chemistry: Introduction to Advanced Electronic Structure Theory}}.
\newblock Courier Corporation, 2012.

\bibitem[Th{\"o}lke and Fabritiis(2022)]{tholke2022equivariant}
Philipp Th{\"o}lke and Gianni~De Fabritiis.
\newblock Equivariant transformers for neural network based molecular potentials.
\newblock In \emph{International Conference on Learning Representations}, 2022.

\bibitem[Thomas et~al.(2018)Thomas, Smidt, Kearnes, Yang, Li, Kohlhoff, and Riley]{thomas2018tensor}
Nathaniel Thomas, Tess Smidt, Steven Kearnes, Lusann Yang, Li~Li, Kai Kohlhoff, and Patrick Riley.
\newblock Tensor field networks: Rotation-and translation-equivariant neural networks for 3{D} point clouds.
\newblock \emph{arXiv Preprint, arXiv:1802.08219}, 2018.

\bibitem[Tran et~al.(2023)Tran, Lan, Shuaibi, Wood, Goyal, Das, Heras-Domingo, Kolluru, Rizvi, Shoghi, et~al.]{oc22}
Richard Tran, Janice Lan, Muhammed Shuaibi, Brandon~M Wood, Siddharth Goyal, Abhishek Das, Javier Heras-Domingo, Adeesh Kolluru, Ammar Rizvi, Nima Shoghi, et~al.
\newblock The open catalyst 2022 {(OC22)} dataset and challenges for oxide electrocatalysts.
\newblock \emph{ACS Catalysis}, 13\penalty0 (5):\penalty0 3066--3084, 2023.

\bibitem[Unke et~al.(2021)Unke, Bogojeski, Gastegger, Geiger, Smidt, and M{\"u}ller]{unke2021se}
Oliver Unke, Mihail Bogojeski, Michael Gastegger, Mario Geiger, Tess Smidt, and Klaus-Robert M{\"u}ller.
\newblock {SE(3)-equivariant prediction of molecular wavefunctions and electronic densities}.
\newblock \emph{Advances in Neural Information Processing Systems}, 34:\penalty0 14434--14447, 2021.

\bibitem[Wang et~al.(2022)Wang, Liu, Lin, Liu, and Ji]{wang2022comenet}
Limei Wang, Yi~Liu, Yuchao Lin, Haoran Liu, and Shuiwang Ji.
\newblock {ComENet}: Towards complete and efficient message passing for {3D} molecular graphs.
\newblock In \emph{The 36th Annual Conference on Neural Information Processing Systems}, pages 650--664, 2022.

\bibitem[Wang et~al.(2009)Wang, Xiao, Suzek, Zhang, Wang, and Bryant]{wang2009pubchem}
Yanli Wang, Jewen Xiao, Tugba~O Suzek, Jian Zhang, Jiyao Wang, and Stephen~H Bryant.
\newblock {PubChem}: a public information system for analyzing bioactivities of small molecules.
\newblock \emph{Nucleic Acids Research}, 37\penalty0 (suppl\_2):\penalty0 W623--W633, 2009.

\bibitem[Wang et~al.(2017)Wang, Bryant, Cheng, Wang, Gindulyte, Shoemaker, Thiessen, He, and Zhang]{wang2017pubchem}
Yanli Wang, Stephen~H Bryant, Tiejun Cheng, Jiyao Wang, Asta Gindulyte, Benjamin~A Shoemaker, Paul~A Thiessen, Siqian He, and Jian Zhang.
\newblock {PubChem} bioassay: 2017 update.
\newblock \emph{Nucleic Acids Research}, 45\penalty0 (D1):\penalty0 D955--D963, 2017.

\bibitem[Weigend and Ahlrichs(2005)]{weigend2005balanced}
Florian Weigend and Reinhart Ahlrichs.
\newblock Balanced basis sets of split valence, triple zeta valence and quadruple zeta valence quality for h to rn: Design and assessment of accuracy.
\newblock \emph{Physical Chemistry Chemical Physics}, 7\penalty0 (18):\penalty0 3297--3305, 2005.

\bibitem[Yan et~al.(2022)Yan, Liu, Lin, and Ji]{yan2022periodic}
Keqiang Yan, Yi~Liu, Yuchao Lin, and Shuiwang Ji.
\newblock Periodic graph transformers for crystal material property prediction.
\newblock In \emph{The 36th Annual Conference on Neural Information Processing Systems}, pages 15066--15080, 2022.

\bibitem[Yu et~al.(2023)Yu, Xu, Qian, Qian, and Ji]{yu2023efficient}
Haiyang Yu, Zhao Xu, Xiaofeng Qian, Xiaoning Qian, and Shuiwang Ji.
\newblock Efficient and equivariant graph networks for predicting quantum {Hamiltonian}.
\newblock In \emph{Proceedings of the 40th International Conference on Machine Learning}, 2023.

\bibitem[Zhang et~al.(2023)Zhang, Wang, Helwig, Luo, Fu, Xie, Liu, Lin, Xu, Yan, Adams, Weiler, Li, Fu, Wang, Yu, Xie, Fu, Strasser, Xu, Liu, Du, Saxton, Ling, Lawrence, St{\"a}rk, Gui, Edwards, Gao, Ladera, Wu, Hofgard, Tehrani, Wang, Daigavane, Bohde, Kurtin, Huang, Phung, Xu, Joshi, Mathis, Azizzadenesheli, Fang, Aspuru-Guzik, Bekkers, Bronstein, Zitnik, Anandkumar, Ermon, Li{\`o}, Yu, G{\"u}nnemann, Leskovec, Ji, Sun, Barzilay, Jaakkola, Coley, Qian, Qian, Smidt, and Ji]{zhang2023artificial}
Xuan Zhang, Limei Wang, Jacob Helwig, Youzhi Luo, Cong Fu, Yaochen Xie, Meng Liu, Yuchao Lin, Zhao Xu, Keqiang Yan, Keir Adams, Maurice Weiler, Xiner Li, Tianfan Fu, Yucheng Wang, Haiyang Yu, YuQing Xie, Xiang Fu, Alex Strasser, Shenglong Xu, Yi~Liu, Yuanqi Du, Alexandra Saxton, Hongyi Ling, Hannah Lawrence, Hannes St{\"a}rk, Shurui Gui, Carl Edwards, Nicholas Gao, Adriana Ladera, Tailin Wu, Elyssa~F. Hofgard, Aria~Mansouri Tehrani, Rui Wang, Ameya Daigavane, Montgomery Bohde, Jerry Kurtin, Qian Huang, Tuong Phung, Minkai Xu, Chaitanya~K. Joshi, Simon~V. Mathis, Kamyar Azizzadenesheli, Ada Fang, Al{\'a}n Aspuru-Guzik, Erik Bekkers, Michael Bronstein, Marinka Zitnik, Anima Anandkumar, Stefano Ermon, Pietro Li{\`o}, Rose Yu, Stephan G{\"u}nnemann, Jure Leskovec, Heng Ji, Jimeng Sun, Regina Barzilay, Tommi Jaakkola, Connor~W. Coley, Xiaoning Qian, Xiaofeng Qian, Tess Smidt, and Shuiwang Ji.
\newblock Artificial intelligence for science in quantum, atomistic, and continuum systems.
\newblock \emph{arXiv preprint arXiv:2307.08423}, 2023.

\end{thebibliography}
\newpage

\appendix
\section{Appendix}
\begin{table}[!h]
    \centering
    \caption{Statistics of the datasets in Training/Validatioin/Testing datasets.}
    \label{tab:statistics-full}
    \resizebox{1\columnwidth}{!}{
    \begin{tabular}{lccccc}
    \toprule 
    Statistic Type &   & QH9-stable-id & QH9-stable-ood &  QH9-dynamic-300k-geo  & QH9-dynamic-300k-mol      \\
    \midrule
    \multirow{4}{*}{\# of atoms} & Mean & 18.02 / 18.02 / 18.03 &  16.97 / 21.25 / 23.67 & 18.04 / 18.04 / 18.04 & 18.01 / 18.15 / 18.11 \\
    & Median    & 18 / 18 / 18 &  17 / 21 / 23 & 18 / 18 / 18  & 18 / 18 / 18 \\
    & Max       & 29 / 29 / 29 &  20 / 22 / 29 & 27 / 27 / 27 & 27 / 25 / 25 \\
    & Min       & 3 / 6 / 4    &   3 / 21 / 23 & 7 / 7 / 7 & 7 / 10 / 9 \\
    \midrule
    \multirow{4}{*}{\# of electronics} & Mean & 65.89 / 65.90 / 65.86 & 64.98 / 68.83 / 70.57 &  65.88 / 65.88 / 65.88 & 65.91 / 65.71 / 65.78 \\
    & Median    & 66 / 66 / 66 & 66 / 70 / 70 & 66 / 66 / 66 & 66 / 66 / 66 \\
    & Max       & 74 / 74 / 74 & 74 / 74 / 74 & 74 / 74 / 74 & 74 / 74 / 72 \\
    & Min       & 10 / 18 / 24 & 10 / 56 / 58 & 24 / 24 /24 & 24 / 34 / 50 \\
    \midrule
    \multirow{4}{*}{Size of Hamiltonian} &  Mean & 141.59 / 141.62 / 141.56 & 138.99 / 149.85 / 155.09 & 141.59 / 141.59 / 141.59 & 141.58 / 141.18 / 142.03 \\
    & Median   & 144 / 144 / 144 & 142 / 150 / 154 & 144 / 144 / 144 & 144 / 144 / 144 \\
    & Max      & 166 / 166 / 166 & 148 / 152 / 166 & 162 / 162 / 162 & 162 / 158 / 158 \\
    & Min      & 18 / 36 / 48    & 18 / 126 / 130 & 54 / 54 / 54 & 54 / 72 / 102 \\
    \midrule 
    \midrule
    Statistic Type &  & QH9-dynamic-100k-geo  & QH9-dynamic-100k-mol & &      \\
    \midrule
    \multirow{4}{*}{\# of atoms} & Mean & 16.53 / 16.53 / 16.53 & 16.56 / 16.38 / 16.38 & & \\
    & Median    & 17 / 17 / 17 & 17 / 17 / 17 & & \\
    & Max       & 19 / 19 / 19 & 19 / 19 / 19 & & \\
    & Min       & 10 / 10 / 10 & 10 / 10 / 10 & & \\
    \midrule
    \multirow{4}{*}{\# of electronics} & Mean  &  64.68 / 64.68 / 64.68 & 64.71 / 64.44 / 64.68 & & \\
    & Median   & 66 / 66 / 66 & 66 / 66 / 66 & & \\
    & Max       & 74 / 74 / 74 & 74 / 70 / 70 & &  \\
    & Min      & 38 / 38 / 38 & 40 / 38 / 46 & & \\
    \midrule
    \multirow{4}{*}{Size of Hamiltonian} &  Mean & 137.96 / 137.96 / 137.96 & 138.04 / 137.34 / 137.88 & & \\
    & Median   & 140 / 140 / 140 & 142 / 142 / 142 & & \\
    & Max     & 146 / 146 /146 & 146 / 146 / 146 & & \\
    & Min      & 82 / 82 / 82 & 86 / 82 / 92 & & \\
    \bottomrule
    \end{tabular}
    }
    \label{tab:statistic_details}
\end{table}

\begin{table}[!h]
    \centering
    \caption{Test performance for the QHNet trained on QH9-stable-id and QH9-stable-ood datasets on the intersection of test sets over various molecule size.}
    \resizebox{1\columnwidth}{!}{
    \begin{tabular}{ccccccccc}
    \toprule 
    \multirow{2}{*}{\# of Atoms} & \multirow{2}{*}{\# of Samples} & \multirow{2}{*}{Training schema} &\multicolumn{3}{c}{$\mathbf{H}$ $[10^{-6} E_h] \downarrow$}  &  \multirow{2}{*}{$\boldsymbol{\epsilon}$ $[10^{-6} E_h] \downarrow$}  & \multirow{2}{*}{$\psi$ $[10^{-2}] \uparrow$} \\
    & & & diag & non-diag & all & & & \\
    \midrule
    \multirow{2}{*}{23}	&  \multirow{2}{*}{620}	& ID	& 86.22 &	57.75&	59.31& 	435.88& 	94.83 \\
    		&& OOD	& 107.09 &	69.87&	71.92& 	583.76& 	93.47 \\
    \midrule
    \multirow{2}{*}{24}	&  \multirow{2}{*}{82}	& ID	& 92.25 &	58.53&	60.31& 	353.93& 	96.92 \\
    		&& OOD	& 125.20 &	75.29&	77.93& 	365.96& 	96.74 \\
    \midrule
    \multirow{2}{*}{25}	&  \multirow{2}{*}{180}	& ID	& 76.73 &	50.99&	52.30& 	505.51& 	96.34 \\
    		&& OOD	& 118.84 &	69.56&	72.07& 	1135.47& 	94.38 \\
    \midrule
    \multirow{2}{*}{26}	&  \multirow{2}{*}{4}	& ID	& 79.75 &	51.49&	52.88& 	254.94& 	97.66 \\
    		&& OOD	& 131.86 &	72.89&	75.79& 	406.69& 	95.73 \\
    \midrule
    \multirow{2}{*}{27}	&  \multirow{2}{*}{33}	& ID	& 69.33 &	46.10&	47.21& 	264.35& 	95.27 \\
    		&& OOD	& 146.63 &	73.43&	76.90& 	491.40& 	91.97 \\
    \midrule
    \multirow{2}{*}{28}	&  \multirow{2}{*}{0} & ID	 &--&--&--&--&--  \\
    		&& OOD 	 &--&--&--&--&--	  \\
    \midrule
    \multirow{2}{*}{29}	&  \multirow{2}{*}{4} &	  ID	& 71.42&  45.87&	47.00&  243.20&	94.55 \\
    		&& OOD	& 223.41 &	90.25&	96.15& 	772.10& 	90.61        \\
    \bottomrule
    \end{tabular}
    }
    \label{tab:performance-over-molecule-size}
\end{table}

\begin{table}[!h]
    \centering
    \caption{Test performance with various training set sizes.} 
    \resizebox{1\columnwidth}{!}{
    \begin{tabular}{ccc|ccccc}
    \toprule
     \multirow{2}{*}{Training sets} & \multirow{2}{*}{\# of examples} & \multirow{2}{*}{Models} & \multicolumn{3}{c}{$\mathbf{H}$ $[10^{-6} E_h] \downarrow$}  &  \multirow{2}{*}{$\boldsymbol{\epsilon}$ $[10^{-6} E_h] \downarrow$}  & \multirow{2}{*}{$\psi$ $[10^{-2}] \uparrow$} \\
     & & & diag & non-diag & all & &  \\
     \midrule
        QH9-Stable-id 	& 104, 664 & QHNet & 111.21 	& 73.68 	& 76.31 	& 798.51 	& 95.85        \\
        QH9-Stable-id-50k 	& 50,000 & QHNet & 116.23 	& 74.26 	& 77.20 	& 1127.33 	& 95.37       \\
        QH9-Stable-id-10k 	& 10,000 & QHNet & 127.07 	& 74.93 	& 78.60 	& 2471.46 	& 96.18       \\
    \bottomrule
    \end{tabular}
    }
    \label{tab:performance-various-training-size}
\end{table}

\subsection{Detailed dataset statistics}
Here we provide a detailed statistics including the number of atoms, the number of electronics and the size of Hamiltonian matrix for training, validation and test sets in Table \ref{tab:statistic_details} for all the proposed datasets.

\subsection{The analysis of the MD trajectory}
\begin{figure}
    \centering
    \begin{subfigure}[b]{0.45\textwidth}
        \centering
        \includegraphics[width=\textwidth]{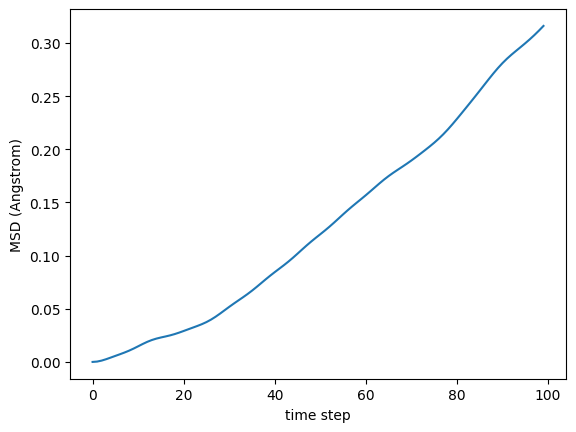}
        \caption{The mean square distance (MSD) between the initial geometry and the geometry in the MD trajectory. Here it demonstrates the first MD trajectory in QH9-dynamic-300k. }
        \label{fig:MSD_trajectory}
    \end{subfigure}
    \hfill
    \begin{subfigure}[b]{0.45\textwidth}
        \centering
        \includegraphics[width=\textwidth]{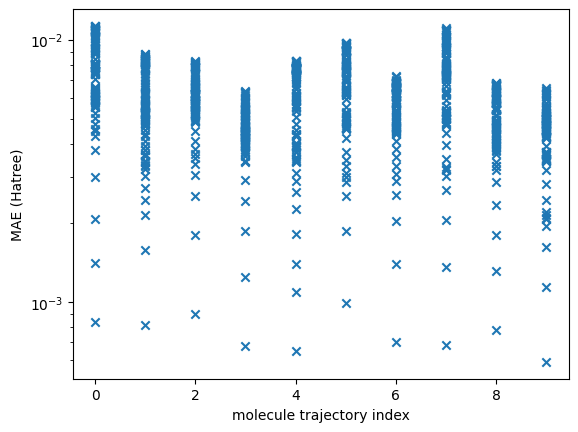}
        \caption{The mean absolute error (MAE) between the Hamiltonian matrix for the initial geometry and the Hamiltonian matrix for the geometry in the MD trajectory in QH9-dynamic-300k.}
        \label{fig:MAE_Ham_traj}
    \end{subfigure}
\end{figure}

To analyze the MD trajectory, we provide the MSD between the geometry in the initial structure and the geometry at each time step in Figure \ref{fig:MSD_trajectory}, and find that molecular geometry is changing over time. Furthermore, in Figure \ref{fig:MAE_Ham_traj}, we showcase the MAE of the Hamiltonian for initial structure and Hamiltonian compared to that for other geometries along the same trajectory. Note that the MAE error of the different geometries is much larger than the prediction error.

\subsection{Test performance over molecule size}
To evaluate the model performance and alleviate the influence of molecular size, we test the QHNet trained on QH9-stable-id and QH9-stable-ood datasets, and demonstrate the test performance on the molecule with various sizes in Table \ref{tab:performance-over-molecule-size}.

\subsection{Test performance over various training sets}
To investigate the model performance with various number of examples in training sets, we train the QHNet on datasets with various sizes. As shown in Table \ref{tab:performance-various-training-size}, we can observe that while the MAE on Hamiltonian matrix and cosine similarity on wavefunction $\psi$ are in similar level with various training sets, the MAE on occupied eigen energies $\epsilon$ becomes much worse when the training sets become smaller.

\end{document}